\documentclass[letterpaper,aps,prl,twocolumn,showpacs,groupedaddress]{revtex4} 

\usepackage{graphicx}  
\usepackage{dcolumn}   
\usepackage{bm}        
\usepackage{amssymb}   

\newcommand{\pp}  {\mbox{$p\bar{p}$}}
\newcommand{\roots}  {\mbox{$\sqrt{s}=1.96$~TeV}}
\newcommand{\zmumu}  {\mbox{$Z/\gamma^*\rightarrow\mu^+ \mu^-$}}
\newcommand{\ztau}  {\mbox{$Z/\gamma^*\rightarrow\tau^+ \tau^-$}}
\newcommand{\z}  {\mbox{$Z/\gamma^*$}}
\newcommand{\Z}  {\mbox{$Z/\gamma^*$}}
\newcommand{\W}  {\mbox{$W$}}
\newcommand{\pt}  {\mbox{$p_{T}$}}
\newcommand{\ptz}  {\mbox{$p_{T}^{Z}$}}
\newcommand{\ptj}  {\mbox{$p_{T}^{\mathrm{jet}}$}}

\newcommand{\rapz}  {\mbox{$y^Z$}}
\newcommand{\rapj}  {\mbox{$y^{\mathrm{jet}}$}}
\newcommand{\herwig}  {{\sc herwig}}
\newcommand{\pythia}  {{\sc pythia}}
\newcommand{\alpgen}  {{\sc alpgen}}
\newcommand{\sherpa}  {{\sc sherpa}}
\newcommand{\mcfm}  {{\sc mcfm}}
\newcommand{\geant}  {{\sc geant}}
\newcommand{\jimmy}  {{\sc jimmy}}
\newcommand{\perugia}  {{Perugia*}}
\newcommand{\cteq}  {{CTEQ}}
\newcommand{\mrst}  {{MRST}}
\newcommand{\mstw}  {{MSTW}}

\newcommand{\dphi}  {\mbox{$\Delta\phi(Z, \mathrm{jet})$}}
\newcommand{\drap}  {\mbox{$\Delta y(Z, \mathrm{jet})$}}

\newcommand{\yboost}  {\mbox{$y_{\text{boost}}(Z+\mathrm{jet})$}}
\newcommand{\adrap}  {\mbox{$|\Delta y(Z, \mathrm{jet})|$}}
\newcommand{\ayb}  {\mbox{$|y_{\text{boost}}(Z+\mathrm{jet})|$}}
\newcommand{\ayboost}  {\mbox{$|y_{\text{boost}}(Z+\mathrm{jet})|$}}

\begin{document}

\hspace{5.2in} \mbox{FERMILAB-PUB-09-373-E}

\title{Measurement of {\boldmath $Z/\gamma^*$}+jet+{\boldmath $X$}\ angular distributions in {\boldmath $p\bar{p}$}~collisions at {\boldmath $\sqrt{s}=1.96$}~TeV
}

%
\author{V.M.~Abazov$^{37}$}
\author{B.~Abbott$^{75}$}
\author{M.~Abolins$^{65}$}
\author{B.S.~Acharya$^{30}$}
\author{M.~Adams$^{51}$}
\author{T.~Adams$^{49}$}
\author{E.~Aguilo$^{6}$}
\author{M.~Ahsan$^{59}$}
\author{G.D.~Alexeev$^{37}$}
\author{G.~Alkhazov$^{41}$}
% ...

\affiliation{\vspace{0.1 in}(The D\O\ Collaboration)\vspace{0.1 in}}
% ...
\affiliation{$^{6}$University of Alberta, Edmonton, Alberta, Canada;
                Simon Fraser University, Burnaby, British Columbia, Canada;
                York University, Toronto, Ontario, Canada and
                McGill University, Montreal, Quebec, Canada}
% ...
\affiliation{$^{30}$Tata Institute of Fundamental Research, Mumbai, India}
% ...
\affiliation{$^{37}$Joint Institute for Nuclear Research, Dubna, Russia}
% ...
\affiliation{$^{41}$Petersburg Nuclear Physics Institute,
                St. Petersburg, Russia}
% ...
\affiliation{$^{49}$Florida State University, Tallahassee, Florida 32306, USA}
% ...
\affiliation{$^{51}$University of Illinois at Chicago,
                Chicago, Illinois 60607, USA}
% ...
\affiliation{$^{59}$Kansas State University, Manhattan, Kansas 66506, USA}
% ...
\affiliation{$^{65}$Michigan State University,
                East Lansing, Michigan 48824, USA}
% ...
\affiliation{$^{75}$University of Oklahoma, Norman, Oklahoma 73019, USA}
% ...
 
\date{July 24, 2009}

\begin{abstract}
We present the first measurements at a hadron collider of differential cross sections for \Z+jet+$X$\ production in \dphi, \adrap\ and \ayb.
Vector boson production in association with jets is an excellent probe of QCD and constitutes the main background to many small cross section processes, such as associated Higgs production.
These measurements are crucial tests of the predictions of perturbative QCD and current event generators, which have varied success in describing the data. 
Using these measurements as inputs in tuning event generators will increase the experimental sensitivity to rare signals.
\end{abstract}


\maketitle

\clearpage

The production of the massive vector bosons \W\ and $Z$\  at hadron colliders, such as the Fermilab Tevatron and CERN Large Hadron Collider, has distinctive features which lead to many applications.
The electron and muon decay modes of these bosons are experimentally rather simple to identify in the complex environment of a hadron collider and typically have very little background.
Reconstruction of the boson kinematics from its decay products provides a unique, colorless probe of the underlying hadron collision and any hadronic recoil to the boson.
The production of high energy jets in association with \W\ and $Z$\ bosons (V+jets) typically lies within the regime of perturbative quantum chromodynamics (pQCD), however such final states are difficult to calculate to higher orders in perturbation theory. 
Several tools have been developed for generating V+jets events, from the parton-shower event generators \pythia~\cite{pythia} and \herwig~\cite{herwig}, to generators that combine tree level matrix element calculations with parton showers, such as \alpgen~\cite{alpgen} (using \pythia\ or \herwig\ for showering and hadronization) and \sherpa~\cite{sherpa}.
Comparisons between these generators show they suffer from significant uncertainties and differ in the predicted kinematics of V+jet production~\cite{mc_paper}. 
Studies of such final states are therefore an excellent testing ground for theoretical predictions, and inputs from measurements are needed to improve these models.
V+jets is also an experimental signature of many other processes with significantly smaller cross sections, such as top quark pair production, associated production of the Higgs boson with a $W$\ or $Z$\ boson, and the decays of particles within many supersymmetric scenarios.
Identifying V+jets final states resulting from such rare processes relies upon a precise understanding of the more copious V+jet production predicted purely by QCD.

Previous measurements at the Tevatron have studied the kinematics of inclusive \Z\ production~\cite{cdf_zpt, d0_zpt, d0_zrap}, of the \z\ in events with at least one jet~\cite{my_zjet}, of the jets in \z\ and \W\ events~\cite{d0_zjet, my_zjet, cdf_zjet, cdf_wjet, henrik_zjet}, and of the production of \Z\ and \W\ in association with heavy flavor~\cite{d0_zb, cdf_zb, d0_wc, cdf_wc}.
In this Letter, we describe the first measurements of the angular correlations between the \Z\ and leading jet in \Z+jet+$X$\ production.
Using the decay mode $\z\rightarrow\mu\mu$, differential \Z+jet+$X$\ cross sections are measured, binned in the azimuthal angle between the \z\ and leading jet, \dphi,
the absolute value of the rapidity~\cite{rapidity} difference between the \z\ and leading jet, \adrap,
and the absolute value of the average rapidity of the \z\ and leading jet, \ayboost.
These differential cross sections are normalized to the measured inclusive \z\ cross section, cancelling many systematic uncertainties.
As for a previous measurement of $\Delta\phi$(jet, jet)\ in inclusive two jet production~\cite{d0_dijet}, the \dphi\ distribution is sensitive to QCD radiation.
In the absence of additional radiation, the \z\ and jet would be produced with equal and opposite transverse momenta, with \dphi$=\pi$.
Further non-collinear radiation at any transverse momentum, \pt, results in \dphi\ deviating from $\pi$, with higher \pt\ radiation giving a smaller \dphi.
The rapidity variables, \drap\ and \yboost, have primary contributions from the relative momenta of the incoming partons in the hard scatter, but again are modified by any additional QCD radiation.
These measurements are therefore excellent tests of the inclusion of QCD radiation in theoretical models, without requiring that more jets be observed in the event.
As a result, the measurements avoid the experimental uncertainties that would be associated with requiring additional jets to be reconstructed, and is also sensitive to jets below detector reconstruction thresholds.

This analysis uses a dataset of \pp\ collisions at \roots,  corresponding to an integrated luminosity of $0.97\pm0.06$~fb$^{-1}$~\cite{d0lumi}\ recorded by the D0 detector between April 2002 and February 2006.
A full description of the D0 detector is available elsewhere~\cite{d0det}, and only the components most relevant to this analysis are described here.
Immediately surrounding the \pp\ interaction region are two tracking detectors: a silicon microstrip tracker and a scintillating fiber tracker, housed inside a solenoidal magnet providing a field of approximately 2~T.
These trackers are used to measure the momenta of charged particles and to reconstruct the primary interaction point in each collision.
Outside the solenoid lies a liquid argon and uranium calorimeter which is split into three sections: a central section extending to $|\eta|<1.1$~\cite{eta} and two forward sections covering $1.4<|\eta|<4$.
Scintillating detectors provides additional energy measurements between the central and forward calorimeters.
Outside the calorimeter lie three layers of muon detectors which are a combination of scintillating pixels and drift tubes.
Between the first and second layers lies a 1.8~T iron toroidal magnet, providing an independent muon momentum measurement used in the trigger system.

Events used in this analysis are selected by at least one of a suite of single-muon triggers. 
Each of these triggers uses fast readout from the muon system scintillators and fiber tracker to initially identify events, then information from the full tracking and muon systems to provide further rejection.
Additional requirements are then applied to obtain a sample of \z\ candidate events. 
Using information from the muon detectors and the tracking system, two muons of opposite charge and \pt$>15$~GeV are required, with a dimuon mass in the range $65<M_{\mu\mu}<115$~GeV.
To reject cosmic rays and poorly reconstructed muons, the muon tracks are required to match the reconstructed primary interaction point both transverse and parallel to the beam direction; the two muon tracks are also required not to be collinear.
Finally, the muons are required to be consistent with the \pp\ bunch crossing time, using timing information from the muon system scintillators.

Jets are reconstructed using the D0 Run~II seeded, iterative mid-point cone algorithm~\cite{d0jets} on clusters of energy deposited in the calorimeter. 
The algorithm is configured with a split-merge fraction of 0.5 and cone radius of $\sqrt{(\Delta\phi)^2 + (\Delta y)^2} = 0.5$.
Shape and quality cuts reject spurious jets caused by electrons, photons or noise in the calorimeter. 
Further corrections are applied for the calorimeter response, instrumental out-of-cone showering effects, and additional energy deposits caused by instrumental noise and pile-up from multiple \pp\ interactions and previous \pp\ bunch crossings.
These corrections are derived by balancing the \pt\ in  $\gamma$\ + jet events, where the $\gamma$\ and jet are opposite in $\phi$.
After corrections, jets with $\pt>20$~GeV are selected.

Further selections are applied to limit the measurement to regions with high detection efficiency and well-understood detector performance: the muons are required to have $|\eta|<1.7$, the primary vertex must lie within 50~cm of the center of the detector along the direction of the beam, and only jets with $|y|<2.8$\ are considered.
A total of 59,336 \zmumu\ candidate events are selected before jet requirements, of which 9,927 contain at least one jet passing all selections.
In events containing more than one jet, the highest \pt\ jet is selected to calculate the angular variables.
Finally, the low \ptz\ region is excluded as at low \ptz, the measurement of the \z\ azimuthal angle is dominated by the experimental resolution of the muon \pt.
Additionally, these events may contain a significant fraction of jets from additional parton interactions, essentially uncorrelated with the \z\ production. 
For these reasons, the measurement is carried out in two kinematic regions: first, with \ptz$>25$~GeV, and then also raising the \ptz\ selection to $>45$~GeV to probe the higher \pt\ region in more detail.
These contain 5900 and 2449 events respectively.

The main source of background in this analysis is muons from semi-leptonic decays in high energy jets or $W$+jet production.
This is reduced to negligible levels by limiting the sum of track \pt\ and the calorimeter energy allowed in a cone around each muon, and avoiding the overlap between muons and jets by requiring angular separation $\Delta R=\sqrt{(\Delta\phi)^2 + (\Delta\eta)^2} > 0.5$.
The remaining contribution of $W$+jets is extracted from data by studying the distribution of muons failing these requirements and extrapolating into the signal region, and is found to be $<0.5$\%\ of the final sample.
Other sources of background (top quark, diboson, and \ztau\ production) are estimated using \pythia\ simulation using \cteq6L1 parton distribution functions (PDFs)~\cite{cteq6l1}, and the ``Tune-A'' underlying event settings~\cite{pythia_tune_a}, passed through a \geant-based~\cite{geant} simulation of the D0 detector, and normalized to higher order theoretical predictions~\cite{mcfm, top_cs}.
A 10\% systematic uncertainty is assigned to this normalization.
Backgrounds are found to be negligible almost everywhere, however the top quark contribution is evenly distributed in \dphi\ and accounts for up to 11\% of the data at small values of \dphi. 
The  estimated background contributions are subtracted from all data bins.

To extract differential cross sections, the measured events must be corrected to the particle level~\cite{particle_level}, accounting for detector resolution, acceptance, and efficiency.
These corrections are derived from simulated \Z+jet events generated with \alpgen\ v2.11 using \cteq6L1 PDFs, and showered using \pythia\ v6.413 with the ``Tune-A'' underlying event settings.
These simulated events are then passed through a \geant-based simulation of the D0 detector, and real data events from random bunch crossings are overlaid to reproduce the effects of multiple $p\bar{p}$\ interactions and detector noise.

Further corrections are applied to the simulation in order to improve the description of the data.
The muon trigger is not simulated; instead the trigger efficiency is measured in data, and parameterized in terms of the geometry of the muon system.
This efficiency is then applied on an event-by-event basis to the simulation, with the average efficiency being approximately 88\%.
The efficiencies of the muon reconstruction and isolation requirements are measured in data and in the simulations, and adjustments of order 3\%\ are applied to the simulation to correct for the differences.
The muon \pt\ resolution is studied by comparing the shape of the \z\ mass peak in data and the simulation, and further smearing on $1/\pt$\ of order 5\%\ is applied to the simulation to reproduce the data.
Jet corrections are studied by measuring the \pt\ balance in back-to-back \z+jet configurations, and further scaling and smearing, which reaches a maximum of 4\%\ at low \pt, is applied to the simulation to match data.
Finally, kinematic variables as produced by the event generator are re-weighted to provide a good simultaneous description of the variables important to this analysis: \ptz, \rapz, \ptj, \rapj, as well as \dphi, \drap and \yboost.
After these corrections, the simulation provides a good description of all measured distributions.

To then correct for detector effects, the corresponding particle level quantities must first be defined.
To minimize dependence on models of particle production, decay and radiation, the definitions are based on the stable particles that enter the detector in the simulation (without identifying the origin of these particles) and correspond as closely as possible to the detector level definitions of muons and jets.
 The \z\ candidate is reconstructed from the highest-\pt\ $\mu^+$\ and $\mu^-$\ with $|y^{\mu}|<1.7$, then requiring $65<M_{\mu\mu}<115$~GeV. 
As for the detector level analysis, two samples with \ptz$>25$~GeV and \ptz$>45$~GeV are selected.
The two muons used to reconstruct the \z\, and any photons in a cone of $\Delta R<0.2$\ around those muons (dominated by QED radiation from the muons), are excluded from the jet reconstruction.
All other stable particles are passed to the D0 Run~II jet algorithm, with the same settings as for detector level jets.
Jets with $|y|<2.8$ are considered, then the highest \pt\ jet is selected and required to have $\pt>20$~GeV.

As the distributions of \drap\ and \yboost\ are symmetric around zero, the absolute value of each is considered in order to increase the statistical precision.
The distributions of \dphi, \adrap, and \ayboost\ are then binned, with the binning determined by a combination of detector resolution considerations, maintaining reasonable numbers of data events in each bin, and maximizing sensitivity to shape differences predicted by different models of \z+jet production.
With the chosen binning, the detector resolution causes little migration between bins (bin purities are generally above 90\%, with the purity defined as the fraction of events in a given particle level bin which are in the same bin at detector level). Purities for \dphi\ drop to around 70~\% in some bins, due to the worse experimental resolution. However, in all cases the main effect to correct is detector acceptance and efficiency.
To do this, the distributions of \dphi, \adrap, and \ayboost\ are populated in the simulation independently for the particle level and detector level selections. 
The ratio of particle level to detector level is then applied to data, with a typical correction factor being approximately 2.2, with some dependence on the variable under consideration.
The statistical uncertainty assigned to the resulting differential cross section corresponds to the detector level statistical uncertainty.
The data are then normalized to the total \z\ production cross section (with no jet or \ptz\ requirements, but the same muon $|y|$\ and dimuon mass requirements) measured with this sample.

Finally, systematic uncertainties are assessed.
Several sources are considered, beginning with the uncertainties on \pt\ resolution, energy scale and detector efficiencies for both muons and jets. 
These corrections are shifted individually up and down one standard deviation, and the ratio of particle level to detector level re-derived.
The difference in the final result is assigned as a systematic uncertainty.
The effects of the various kinematic re-weightings are assessed by turning off each of these in turn and repeating the analysis; however, due to the small effects of detector resolution, the results are largely insensitive to these tests except for \dphi, where the lower bin purities lead to a larger correlation between the shape of the distribution and the detector to particle-level correction factor.
The region of low \dphi\ is found to be particularly sensitive to jets from additional interactions in an event, so a further uncertainty is assessed by varying the real data from random bunch crossings overlaid on the simulation.
The systematic uncertainties are combined in quadrature, with the jet energy scale uncertainty generally being the largest single source for \adrap\ and \ayb,  and accounting for approximately half of the total uncertainty.
For \dphi, the uncertainty on the kinematic re-weightings is comparable to or larger than the jet energy scale uncertainty.
For the selection with \ptz$>25$~GeV, the total systematic uncertainty is of comparable size to the statistical uncertainty; for the selection with \ptz$>45$~GeV, the statistical uncertainty dominates.

Predictions for \dphi, \adrap\ and \ayboost\ distributions are obtained from several theoretical models, as well as leading order (LO) and next-to-leading order (NLO) pQCD calculations.
Predictions at LO and NLO are obtained with \mcfm\ v5.6~\cite{mcfm}, together with the \mstw2008 LO and NLO PDFs~\cite{mstw} respectively. 
The distribution at \dphi=$\pi$\ contains divergences and is excluded from comparisons to data.
Re-normalization and factorization scales are set to the sum in quadrature of the mass and \pt\ of the \Z\ in each event, and the dependence on this choice is assessed by varying both scales simultaneously up and down by a factor of two, both for the differential distribution and the inclusive \z\ cross section used in normalization.
PDF uncertainties are assessed using the \mstw2008 68\%\ error sets, again taking into account the effect on the differential distribution and the inclusive \z\ cross section used in normalization. 
These are found to be approximately a factor of two smaller than the scale uncertainties.
Comparisons with data are performed after correcting the parton level prediction from \mcfm\ for the effects of hadronization and the underlying event.
These corrections have been derived from a sample of \Z+jet events generated with \pythia\ v6.421~\cite{pythia_64} using the underlying event tune QW~\cite{pythia_tune} with the \cteq6.1M PDFs~\cite{cteq61}.
They are derived by comparing the full prediction (taken from the final state particles, including the underlying event) to the purely perturbative part (calculated from partons taken after the parton shower, with no underlying event), and are typically around 4\%.
However, the low \dphi\ ($<1.5$~rad) region is dominated by non-perturbative effects, so the pQCD calculation for this bin is excluded.
Corrections for quantum electrodynamic final state radiation (FSR)  from the muons are also derived from the same \pythia\ sample, by comparing the prediction calculated using the muons after FSR to those using the generated boson.
These are typically less than 1\% after accounting for the effect on the inclusive \z\ cross section and are primarily the result of events migrating out of the mass window.

Predictions are also obtained from four current event generators. 
When considering the number of generators available and the various tunes of those generators, a complete survey of the field would be impossible.
We choose to focus on the current matrix element calculations with matched parton showers, as implemented in \sherpa\ and \alpgen, as previous measurements indicate these provide the best description of boson+jets final states~\cite{my_zjet, henrik_zjet}.
The \alpgen\ matrix element calculation can be interfaced to the \pythia\ or \herwig\ parton shower and hadronization models, and we test both.
Further, to assess the impact of the additional matrix elements in \alpgen, we also run \pythia\ and \herwig\ stand-alone.
Unless stated otherwise, we use the same PDF set throughout: CTEQ6.1M. 
First, a sample of events is generated with \sherpa\ v1.1.3 with up to three partons in the matrix element calculation. Parton jets from the matrix element are required to have \pt$> 13$~GeV and $\Delta R ($jet, jet$) > 0.4$.
In \sherpa, both the renormalization and factorization scales are set according to the CKKW prescription~\cite{CKKW}.
To provide a typical uncertainty on a prediction from an event generator, \sherpa\ samples are also generated with the renormalization and factorization scales varied up and down by a factor of two, both for the differential distribution and the inclusive \z\ cross section used for normalization.
A sample of events is then generated with \alpgen\ v2.13, again with up to three partons in the matrix element calculation. 
The factorization scale is set to the sum in quadrature of the mass and \pt\ of the \Z, and the renormalization scale set according to the CKKW prescription.
Parton jets from the matrix element calculation are required to have \pt$> 13$~GeV, and $\Delta R ($jet, jet$) > 0.4$.
These events are hadronized in three ways. First,  using  \herwig\ v6.510 (with \jimmy\ v4.31~\cite{jimmy} for multiple parton interactions).
Then using \pythia\ v6.421 with underlying event tune QW (using the $Q^2$-ordered shower) and the 2-loop prescription for $\alpha_s$.
Finally, using \pythia\ v6.421 with the \pt-ordered shower~\cite{pt_ordered}, for which there is currently no tune using the CTEQ6.1M PDFs, so instead the \perugia\ tune~\cite{perugia} using the \mrst2007 modified LO (LO*) PDFs~\cite{mrstMLO} is used.
This results in three different \alpgen\ predictions, and in each case the default matching procedure is applied after hadronization, with each parton jet required to match a particle level jet with $\pt>18$~GeV, by requiring $\Delta R ($jet, jet$) < 0.4$. 
To determine the impact of the matching to the \alpgen\ matrix elements calculation, \herwig\ and \pythia\ are also run stand-alone to produce three inclusive \zmumu\ samples.
First using \herwig\ v6.510 with \jimmy\ v4.31  for multiple parton interactions.
Then two using \pythia\ v6.421: one with tune QW and the 2-loop prescription for $\alpha_s$; the other with the \perugia\ tune and the \mrst2007 LO* PDFs.
In both \pythia\ and \herwig, the renormalization and factorization scales for the hard scatter are set to the mass of the \z, and for the initial and final state showers are determined dynamically.
For all generators, the particle level quantities are extracted as defined earlier.

\begin{figure*}[!htb]\center
\includegraphics[width=160mm]{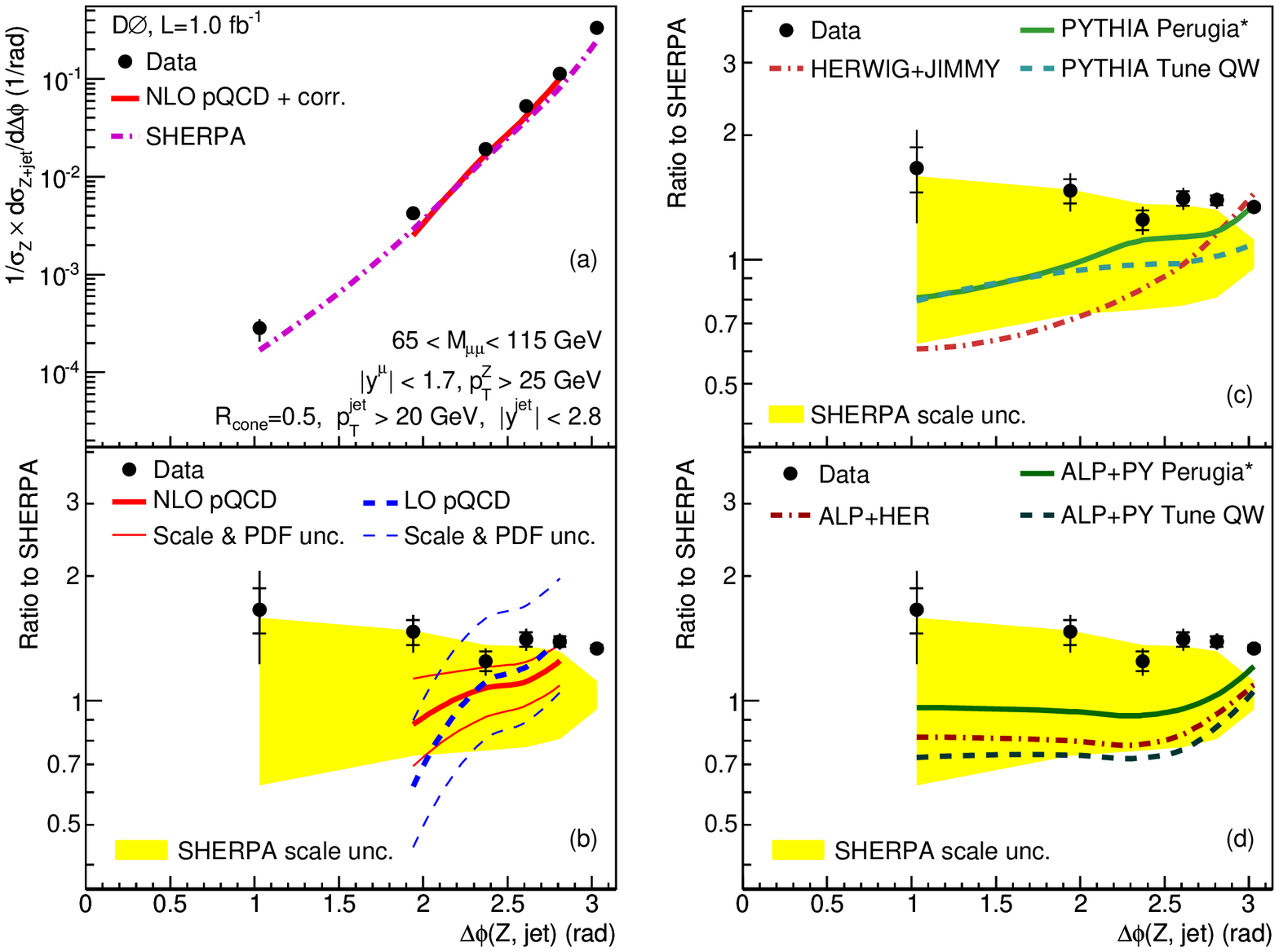}
\caption{\label{fig:dphi_result1}The measured normalized cross section in bins of \dphi\ for \Z+jet+$X$\ events  for \ptz$>25$~GeV. The distribution is shown in (a) and compared to fixed order calculations in (b), parton shower generators in (c), and the same parton shower generators matched to \alpgen\ matrix elements in (d). All ratios in (b), (c), and (d) are shown relative to \sherpa, which provides the best description of data overall.
}
\end{figure*}

\begin{figure*}[!htb]\center
\includegraphics[width=160mm]{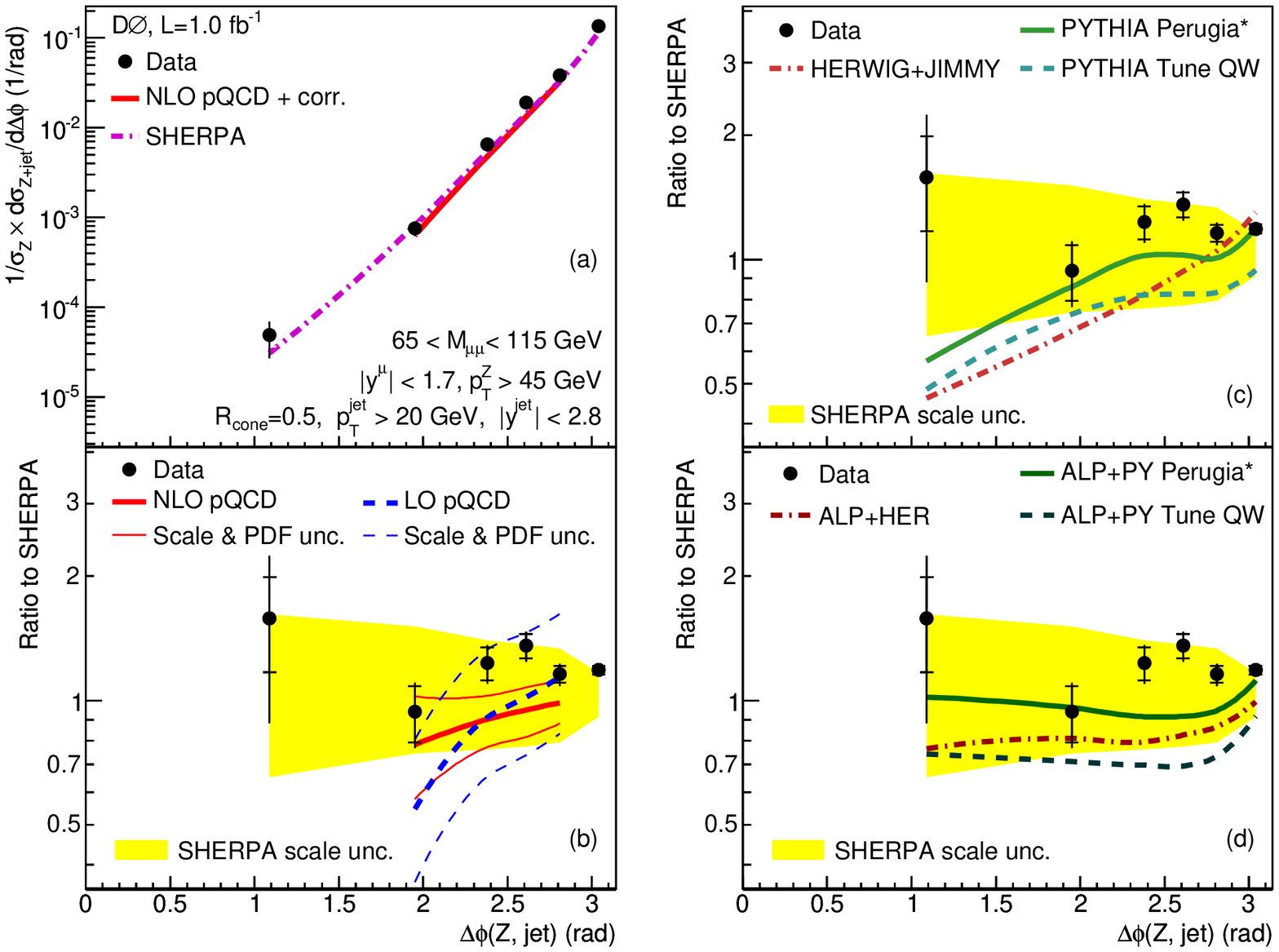}
\caption{\label{fig:dphi_result2}The measured normalized cross section in bins of \dphi\ for \Z+jet+$X$\ events  for \ptz$>45$~GeV. The distribution is shown in (a) and compared to fixed order calculations in (b), parton shower generators in (c), and the same parton shower generators matched to \alpgen\ matrix elements in (d). All ratios in (b), (c), and (d) are shown relative to \sherpa, which provides the best description of data overall.
}
\end{figure*}

\begin{figure*}[!htb]\center
\includegraphics[width=160mm]{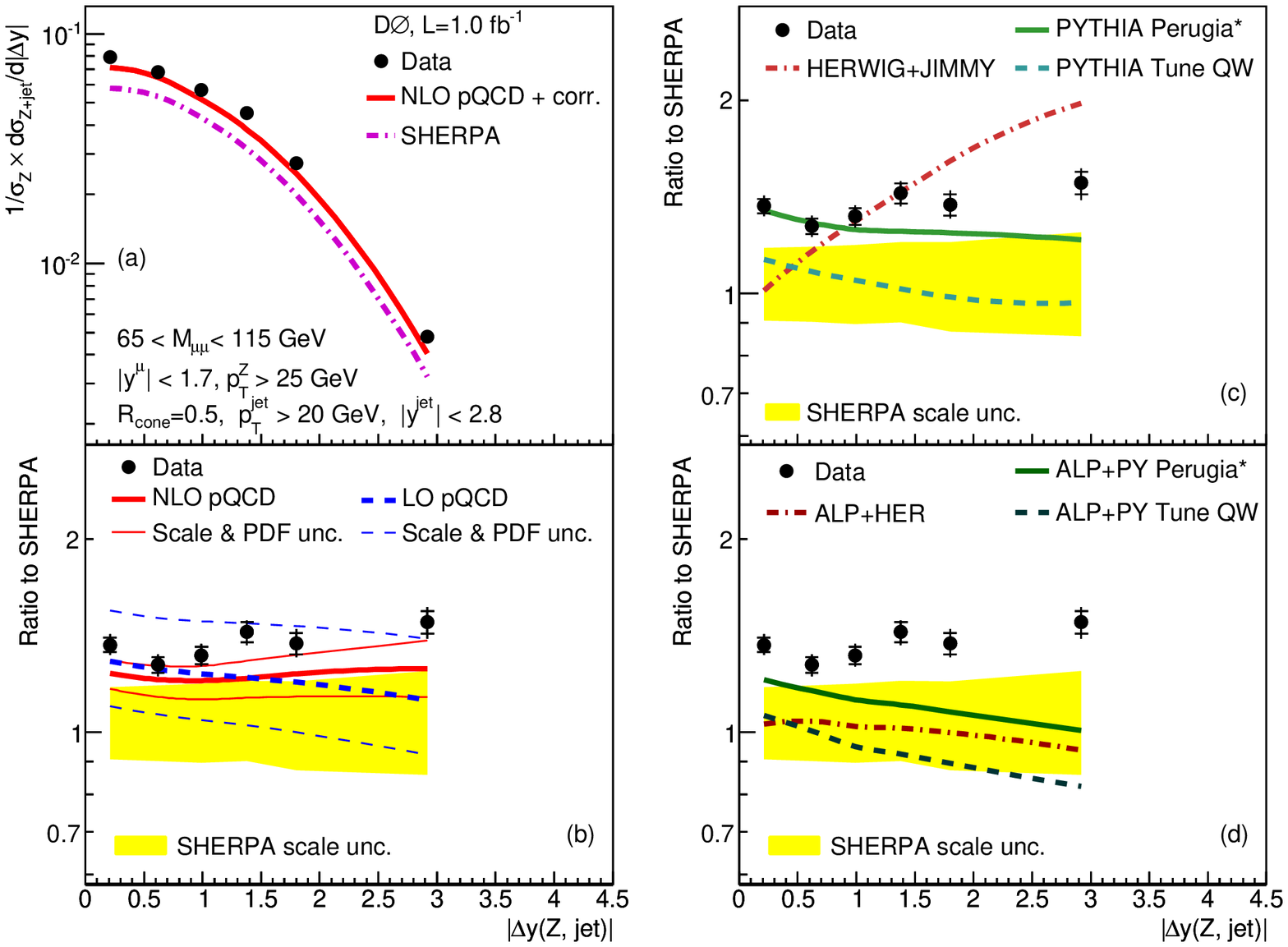}
\caption{\label{fig:drap_result1}The measured normalized cross section in bins of \adrap\ for \Z+jet+$X$\ events  for \ptz$>25$~GeV. The distribution is shown in (a) and compared to fixed order calculations in (b), parton shower generators in (c), and the same parton shower generators matched to \alpgen\ matrix elements in (d). All ratios in (b), (c), and (d) are shown relative to \sherpa, which provides the best description of data overall.
}
\end{figure*}

\begin{figure*}[!htb]\center
\includegraphics[width=160mm]{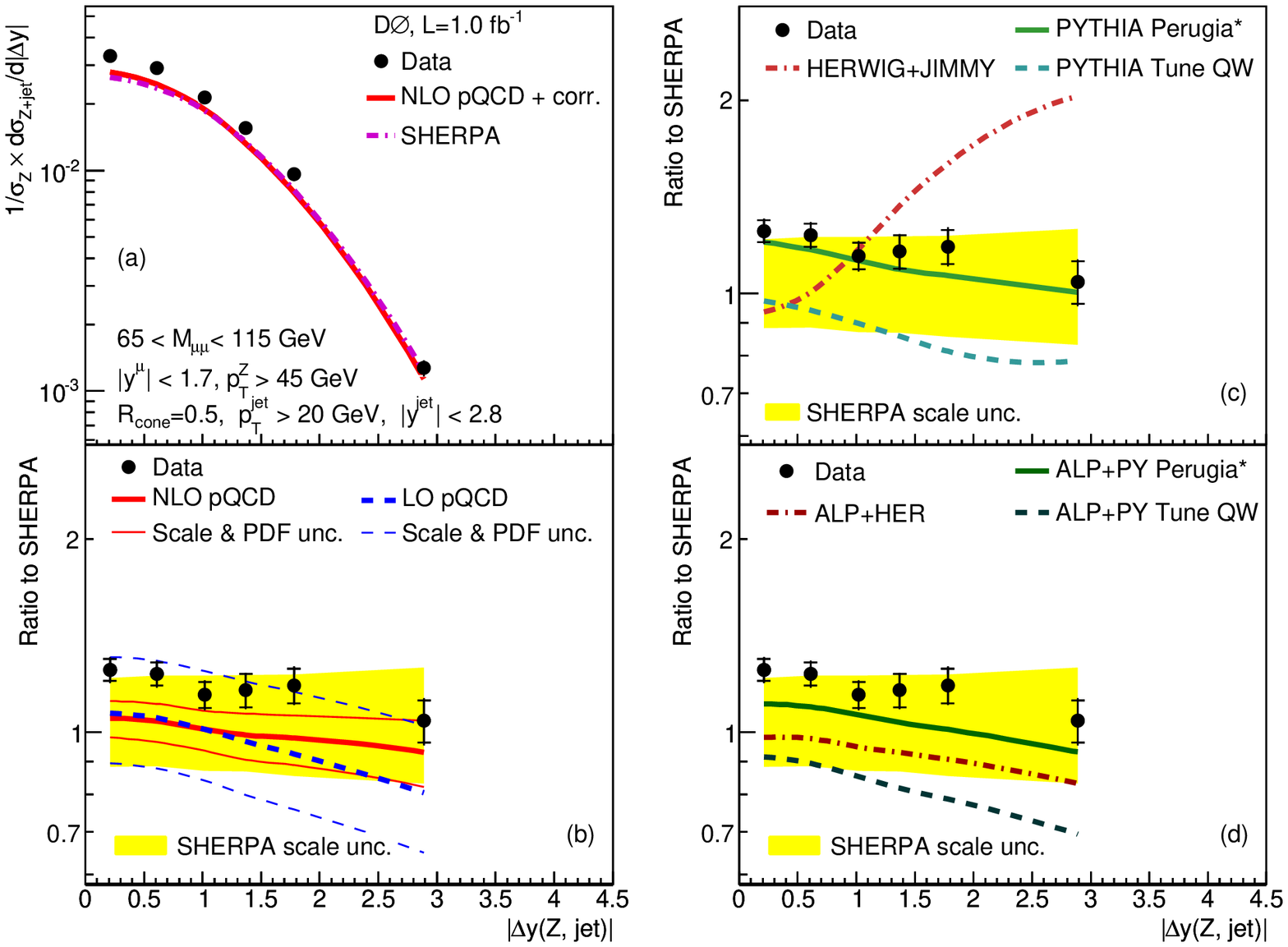}
\caption{\label{fig:drap_result2}The measured normalized cross section in bins of \adrap\ for \Z+jet+$X$\ events  for \ptz$>45$~GeV. The distribution is shown in (a) and compared to fixed order calculations in (b), parton shower generators in (c), and the same parton shower generators matched to \alpgen\ matrix elements in (d). All ratios in (b), (c), and (d) are shown relative to \sherpa, which provides the best description of data overall.
}
\end{figure*}

\begin{figure*}[!htb]\center
\includegraphics[width=160mm]{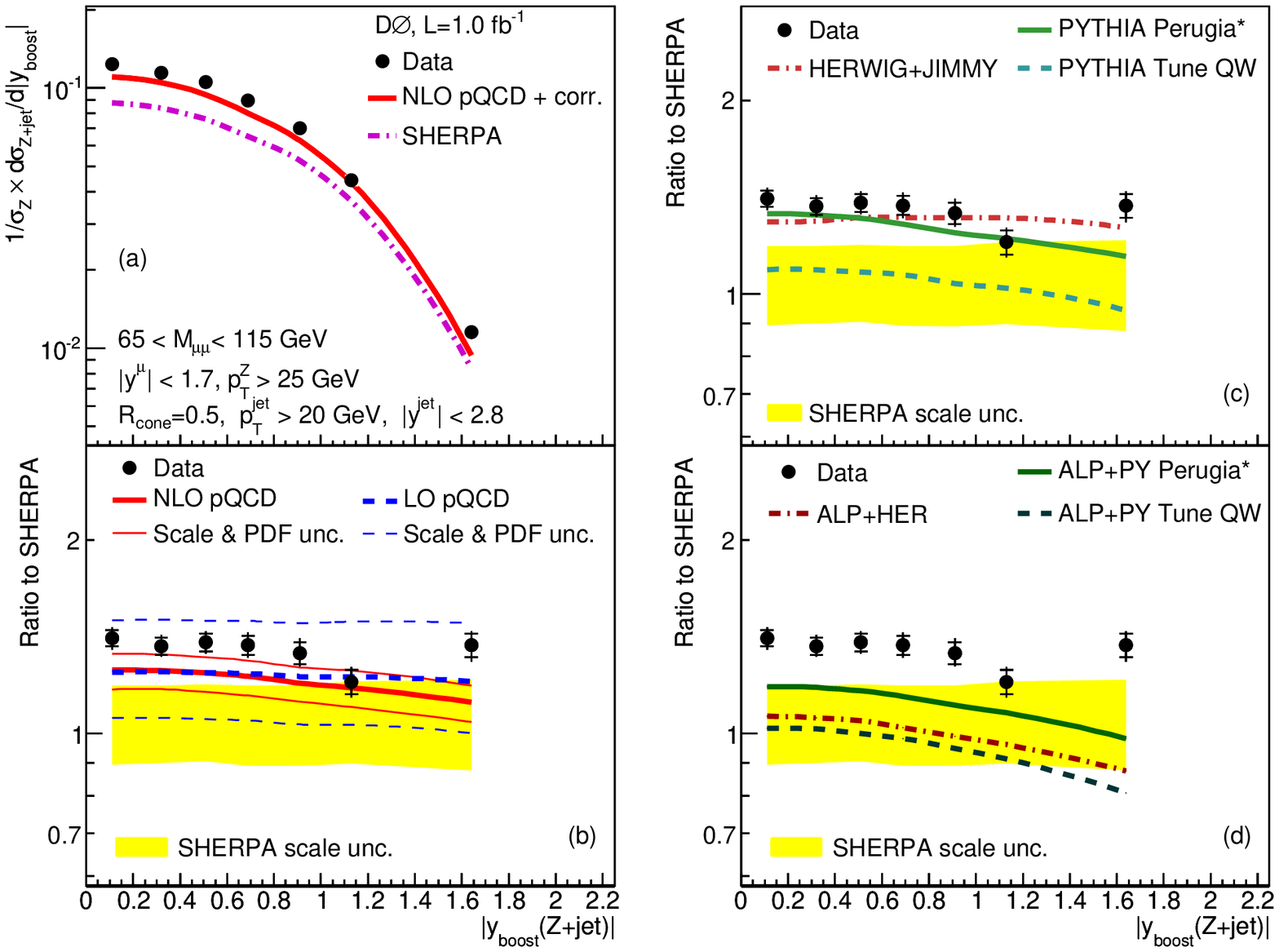}
\caption{\label{fig:yboost_result1}The measured normalized cross section in bins of \ayboost\ for \Z+jet+$X$\ events  for \ptz$>25$~GeV. The distribution is shown in (a) and compared to fixed order calculations in (b), parton shower generators in (c), and the same parton shower generators matched to \alpgen\ matrix elements in (d). All ratios in (b), (c), and (d) are shown relative to \sherpa, which provides the best description of data overall.
}
\end{figure*}

\begin{figure*}[!htb]\center
\includegraphics[width=160mm]{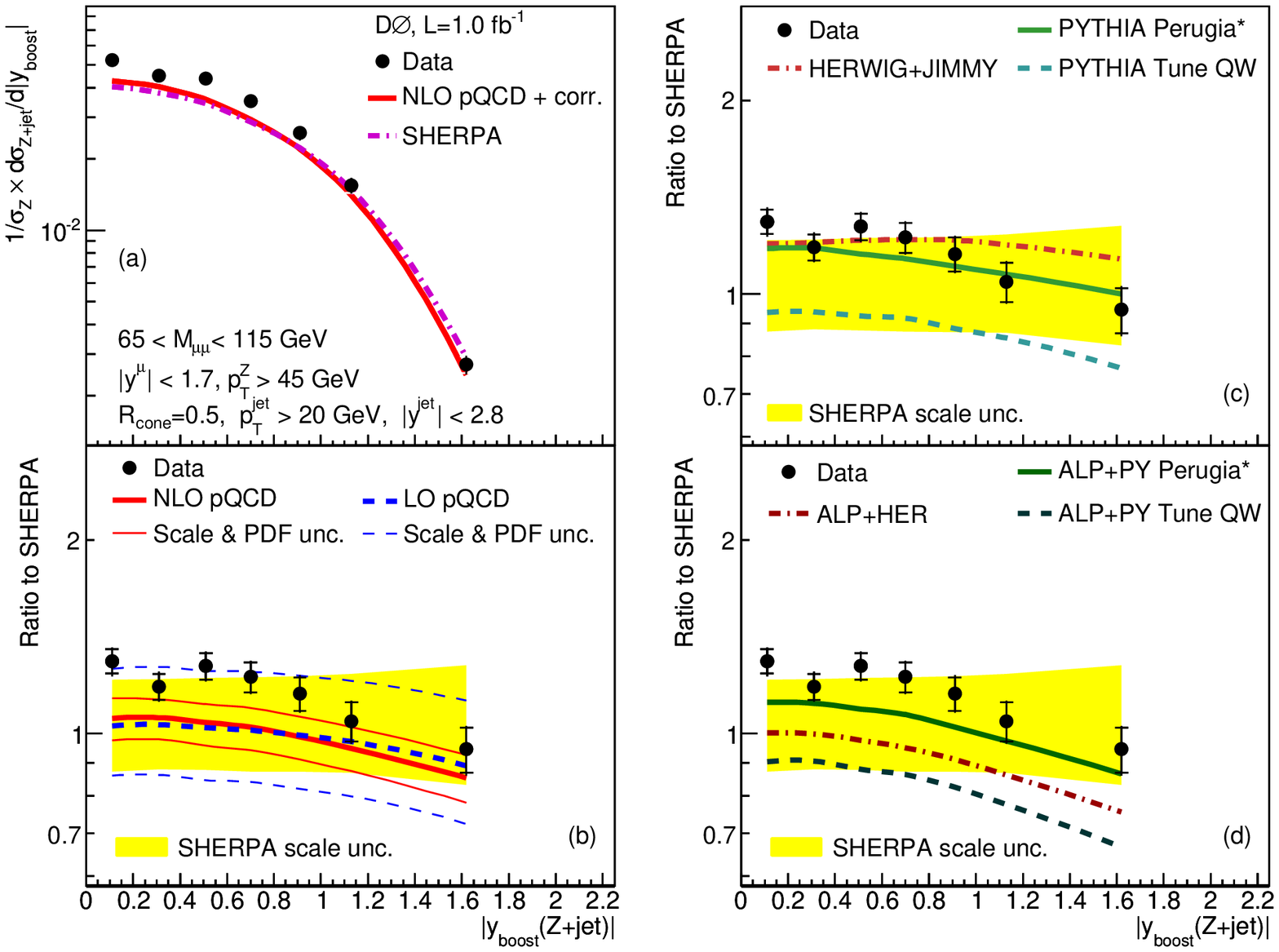}
\caption{\label{fig:yboost_result2}The measured normalized cross section in bins of \ayboost\ for \Z+jet+$X$\ events  for \ptz$>45$~GeV. The distribution is shown in (a) and compared to fixed order calculations in (b), parton shower generators in (c), and the same parton shower generators matched to \alpgen\ matrix elements in (d). All ratios in (b), (c), and (d) are shown relative to \sherpa, which provides the best description of data overall.
}
\end{figure*}

The normalized differential cross sections are available in Ref.~\cite{supplemental}, and presented binned in \dphi\ (Figs. \ref{fig:dphi_result1} and \ref{fig:dphi_result2}, Tables \ref{tab:dphi25} and \ref{tab:dphi45}), \adrap\ (Figs. \ref{fig:drap_result1} and \ref{fig:drap_result2}, Tables \ref{tab:drap25} and \ref{tab:drap45}) and \ayb\ (Figs. \ref{fig:yboost_result1} and \ref{fig:yboost_result2}, Tables \ref{tab:yboost25} and \ref{tab:yboost45}).
The data points are placed at the bin average, defined as point where the differential cross section within the bin, taken from simulation re-weighted to match the shape in data, is equal to the measured value in the bin~\cite{bins}.
The data are shown with statistical uncertainties (inner error bar) and combined statistical and systematic uncertainties (outer error bar). 
For clarity, only the predictions of NLO pQCD and \sherpa\ are shown with the data in part (a) of each figure.
In the other parts of each figure, ratios are shown, where the data and all other theory predictions are divided through by the prediction from \sherpa.
We choose to show the data only in one ratio, to avoid repeating the data uncertainties and statistical fluctuations several times.
\sherpa\ is chosen as the common denominator for all ratios as it provides the best description of the shape of the data in most distributions, simplifying the determination of trends in other theoretical predictions relative to the data.
The \sherpa\ scale uncertainty is shown as a shaded band around unity.

The integrated cross sections are also extracted: $\sigma_{Z+\rm{jet}} / \sigma_{Z}$\ for the stated \z\ and jet selections. These are measured to be 
$[122\pm2(\rm{stat.})\pm4(\rm{syst.})]\times10^{-3}$\ for $\ptz>25$~GeV, and 
$[47\pm1(\rm{stat})\pm2(\rm{syst})]\times10^{-3}$\  for \ptz$>45$~GeV.
In both cases, the \ptz\ requirement is only made on the \z+jet selection, not the inclusive \z\ selection.
The corresponding results from pQCD are 
$[111\pm6(\rm{scale})\pm2(\rm{PDF})]\times10^{-3}$\ at NLO and  
$[112\pm20(\rm{scale})\pm1(\rm{PDF})]\times10^{-3}$\ at LO for \ptz$>25$~GeV, and 
$[40\pm3(\rm{scale})\pm1(\rm{PDF})]\times10^{-3}$\ at NLO and  
$[40\pm8(\rm{scale})\pm1(\rm{PDF})]\times10^{-3}$\ at LO for \ptz$>45$~GeV.

\begin{table}[!htb]
\caption{\label{tab:dphi25}The measured cross section in bins of \dphi\ for \Z\ +jet+$X$\ events with \ptz$>25$~GeV, normalized to the measured \z\ cross section.
}
\begin{ruledtabular}
\begin{tabular}{D{,}{\,\mbox{--}\,}{-1}ccD{,}{\,.\,}{-1}D{,}{\,.\,}{-1}D{,}{\,.\,}{-1}}
\multicolumn{1}{c}\mbox{$\Delta\phi$~~~} & 
\mbox{$\langle \Delta\phi\rangle$ } & 
\mbox{$1/\sigma\times d\sigma/d\Delta\phi$} & 
\multicolumn{1}{c} \mbox{$\delta\sigma_{\rm{stat.}}$}  &
\multicolumn{2}{c} \mbox{$\delta\sigma_{\rm{total}}$~~~~~~~~~} \\

\multicolumn{1}{c}\mbox{(rad)~~} &  
\mbox{(rad)} &  
\mbox{(1/rad)}   & 
\multicolumn{1}{c} \mbox{(\%)~~}   &  
\multicolumn{2}{c} \mbox{(\%)~~~~~~~~~~}  \\
    \hline
0.0 , 1.5 & 1.09 & 2.82 $\times~10^{-4}$ & 12 & +24  & -26  \\
1.5 , 2.2 & 1.95 & 4.22 $\times~10^{-3}$ & 6,9& +10  & -11  \\
2.2 , 2.5 & 2.38 & 1.93 $\times~10^{-2}$ & 5,5& +7,3 & -7,9 \\
2.5 , 2.7 & 2.61 & 5.27 $\times~10^{-2}$ & 4,1& +6,1 & -6,2 \\
2.7 , 2.9 & 2.81 & 1.13 $\times~10^{-1}$ & 2,8& +4,5 & -4,6 \\
2.9 , \pi & 3.04 & 3.32 $\times~10^{-1}$ & 1,7& +3,6 & -3,3 \\

\end{tabular}
\end{ruledtabular}
\end{table}

\begin{table}[!htb]
\caption{\label{tab:dphi45}The measured cross section in bins of \dphi\ for \Z\ +jet+$X$\ events with \ptz$>45$~GeV, normalized to the measured \z\ cross section.
}
\begin{ruledtabular}
\begin{tabular}{D{,}{\,\mbox{--}\,}{-1}ccD{,}{\,.\,}{-1}D{,}{\,.\,}{-1}D{,}{\,.\,}{-1}}
\multicolumn{1}{c}\mbox{$\Delta\phi$~~~} & 
\mbox{$\langle \Delta\phi\rangle$ } & 
\mbox{$1/\sigma\times d\sigma/d\Delta\phi$} & 
\multicolumn{1}{c} \mbox{$\delta\sigma_{\rm{stat.}}$}  &
\multicolumn{2}{c} \mbox{$\delta\sigma_{\rm{total}}$~~~~~~~~~} \\

\multicolumn{1}{c}\mbox{(rad)~~} &  
\mbox{(rad)} &  
\mbox{(1/rad)}   & 
\multicolumn{1}{c} \mbox{(\%)~~}   &  
\multicolumn{2}{c} \mbox{(\%)~~~~~~~~~~}  \\
    \hline
0.0 , 1.5 & 1.09 & 4.88 $\times~10^{-5}$ & 26 & +42 & -44 \\
1.5 , 2.2 & 1.95 & 7.61 $\times~10^{-4}$ & 16 & +18 & -18 \\
2.2 , 2.5 & 2.38 & 6.57 $\times~10^{-3}$ & 9,1 & +11 & -11 \\
2.5 , 2.7 & 2.61 & 1.91 $\times~10^{-2}$ & 6,9 & +8,2 & -8,6 \\
2.7 , 2.9 & 2.81 & 3.83 $\times~10^{-2}$ & 4,8 & +6,0 & -6,5 \\
2.9 , \pi & 3.04 & 1.35 $\times~10^{-1}$ & 2,5 & +3,5 & -3,5 \\
\end{tabular}
\end{ruledtabular}
\end{table}

\begin{table}[!htb]
\caption{\label{tab:drap25}The measured cross section in bins of \adrap\ for \Z+jet+$X$\ events with \ptz$>25$~GeV, normalized to the measured \z\ cross section.
}
\begin{ruledtabular}
\begin{tabular}{ccccc}
$|\Delta y|$ & $\langle |\Delta y|\rangle$ & $1/\sigma\times d\sigma/d|\Delta y|$ & $\delta\sigma_{\rm{stat.}}$  & $\delta\sigma_{\rm{total}}$ \\
 &  &    & (\%)  & (\%)  \\
    \hline
0.00 -- 0.40 & 0.21 & 7.91 $\times~10^{-2}$ & 2.6 & +3.9 $-3.6$ \\
0.40 -- 0.80 & 0.61 & 6.79 $\times~10^{-2}$ & 2.8 & +3.8 $-3.9$ \\
0.80 -- 1.20 & 1.02 & 5.68 $\times~10^{-2}$ & 3.0 & +4.1 $-4.0$ \\
1.20 -- 1.55 & 1.37 & 4.52 $\times~10^{-2}$ & 3.6 & +4.3 $-4.6$ \\
1.55 -- 2.05 & 1.78 & 2.74 $\times~10^{-2}$ & 3.8 & +5.4 $-4.8$ \\
2.05 -- 4.50 & 2.89 & 4.80 $\times~10^{-3}$ & 4.0 & +5.5 $-5.8$ \\
\end{tabular}
\end{ruledtabular}
\end{table}

\begin{table}[!htb]
\caption{\label{tab:drap45}The measured cross section in bins of \adrap\ for \Z+jet+$X$\ events with \ptz$>45$~GeV, normalized to the measured \z\ cross section.
}
\begin{ruledtabular}
\begin{tabular}{ccccc}
$|\Delta y|$ & $\langle |\Delta y|\rangle$ & $1/\sigma\times d\sigma/d|\Delta y|$ & $\delta\sigma_{\rm{stat.}}$  & $\delta\sigma_{\rm{total}}$ \\
 &  &    & (\%)  & (\%)  \\
    \hline
0.00 -- 0.40 & 0.21 & 3.31 $\times~10^{-2}$ & 3.9 & +4.6 $-4.6$ \\
0.40 -- 0.80 & 0.61 & 2.91 $\times~10^{-2}$ & 4.1 & +4.9 $-4.9$ \\
0.80 -- 1.20 & 1.02 & 2.14 $\times~10^{-2}$ & 4.7 & +5.3 $-5.5$ \\
1.20 -- 1.55 & 1.37 & 1.56 $\times~10^{-2}$ & 5.9 & +6.5 $-6.5$ \\
1.55 -- 2.05 & 1.78 & 9.60 $\times~10^{-3}$ & 6.2 & +7.1 $-6.8$ \\
2.05 -- 4.50 & 2.89 & 1.27 $\times~10^{-3}$ & 7.6 & +8.3 $-8.5$ \\
\end{tabular}
\end{ruledtabular}
\end{table}

\begin{table}[!htb]
\caption{\label{tab:yboost25}The measured cross section in bins of \ayb\ (denoted $|y_{\text{b}}|$) for \Z+jet+$X$\ events with \ptz$>25$~GeV, normalized to the measured \z\ cross section.
}
\begin{ruledtabular}
\begin{tabular}{ccccc}
$|y_{\text{b}}|$ & $\langle |y_{\text{b}}|\rangle$ & $1/\sigma\times d\sigma/d|y_{\text{b}}|$ & $\delta\sigma_{\rm{stat.}}$  & $\delta\sigma_{\rm{total}}$ \\
 & &    & (\%)  & (\%)  \\
    \hline
0.00 -- 0.20 & 0.11 & 1.24 $\times~10^{-1}$ & 2.9 & +4.0 $-4.0$ \\
0.20 -- 0.40 & 0.31 & 1.15 $\times~10^{-1}$ & 3.0 & +4.2 $-3.8 $ \\
0.40 -- 0.60 & 0.51 & 1.05 $\times~10^{-1}$ & 3.2 & +4.1 $-4.0 $ \\
0.60 -- 0.80 & 0.70 & 8.95 $\times~10^{-2}$ & 3.4 & +4.7 $-4.5 $ \\
0.80 -- 1.00 & 0.91 & 7.01 $\times~10^{-2}$ & 3.8 & +5.4 $-4.8 $ \\
1.00 -- 1.25 & 1.13 & 4.42 $\times~10^{-2}$ & 4.3 & +5.4 $-5.4 $ \\
1.25 -- 2.25 & 1.62 & 1.15 $\times~10^{-2}$ & 4.2 & +5.3 $-5.8 $ \\
\end{tabular}
\end{ruledtabular}
\end{table}

\begin{table}[!htb]
\caption{\label{tab:yboost45}The measured cross section in bins of \ayb\ (denoted $|y_{\text{b}}|$) for \Z+jet+$X$\ events with \ptz$>45$~GeV, normalized to the measured \z\ cross section.
}
\begin{ruledtabular}
\begin{tabular}{ccccc}

$|y_{\text{b}}|$ & $\langle |y_{\text{b}}|\rangle$ & $1/\sigma\times d\sigma/d|y_{\text{b}}|$ & $\delta\sigma_{\rm{stat.}}$  & $\delta\sigma_{\rm{total}}$ \\
 & &    & (\%)  & (\%)  \\
    \hline
0.00 -- 0.20 & 0.11 & 5.23 $\times~10^{-2}$ & 4.3 & +5.2 $-5.2$ \\
0.20 -- 0.40 & 0.31 & 4.50 $\times~10^{-2}$ & 4.6 & +5.4 $-5.3$ \\
0.40 -- 0.60 & 0.51 & 4.36 $\times~10^{-2}$ & 4.7 & +5.4 $-5.5$ \\
0.60 -- 0.80 & 0.70 & 3.50 $\times~10^{-2}$ & 5.3 & +6.1 $-6.0$ \\
0.80 -- 1.00 & 0.91 & 2.57 $\times~10^{-2}$ & 6.1 & +6.7 $-6.6$ \\
1.00 -- 1.25 & 1.13 & 1.55 $\times~10^{-2}$ & 7.0 & +7.6 $-7.7$ \\
1.25 -- 2.25 & 1.62 & 2.72 $\times~10^{-3}$ & 8.1 & +8.8 $-9.1$ \\
\end{tabular}

\end{ruledtabular}
\end{table}

Where it is valid, the NLO pQCD calculation provides a good description of the data and is a significant improvement in both shape and uncertainty over LO.
However, an overall normalization difference of just over 1 standard deviation of the combined data and NLO theoretical uncertainties is observed, and slightly larger in the \ptz$>45$~GeV sample.
Of the event generators, \herwig\ shows significant disagreement with data in both \dphi\ and \adrap. 
The trend in the \adrap\ is consistent with the description of the leading jet rapidity~\cite{my_zjet} by \herwig, and is significantly improved when interfaced to the \alpgen\ matrix element calculation.
The predictions of the new \perugia\ tune of \pythia\ provide a good description of the data in \adrap\ and \ayb, but not in \dphi.
The modelling of \dphi\ is improved when \pythia\ is interfaced to \alpgen.
In general, the three predictions obtained from \alpgen\ provide a good description of the shape of \adrap\ and \ayb\ for \ptz$>45$~GeV, but perform less well for the other distributions measured.
Overall, \sherpa\  provides the best description of the shape of data, but shows a significant normalization difference.
Further, in the sample with \ptz$>45$~GeV the \sherpa\ description of \ayb\ shows a slope relative to the data, which may also be present in \adrap\ though it is less clear.
All event generators suffer from significant scale uncertainties, of comparable size to the uncertainty on the LO pQCD prediction. 
For \adrap\ and \ayb, these uncertainties change the normalization with little effect on the shape. 
In the case of \dphi, there is some shape dependence, indicating that \alpgen\ could be brought into better agreement with the data through a lower scale choice. 
Studies carried out by shifting the renormalization and factorization scales down in \alpgen\ and the corresponding \pythia\ showering confirm that this is the case.

In summary, we have presented the first measurements at a hadron collider of the \Z+jet+$X$\ normalized differential cross section in \dphi, \adrap, and \ayboost.
The measurements were made using a sample corresponding to  $0.97\pm0.06$~fb$^{-1}$\ of integrated luminosity recorded by the D0 experiment in \pp\ collisions at \roots.
These measurement test the current best predictions for vector boson + jet production at hadron colliders, and are essential inputs for the tuning of event generators.
Improving the modeling of this important signal will lead to increased sensitivity of searches for rare and new physics.

%
We thank John Campbell for useful discussions and input on \mcfm.
We thank the staffs at Fermilab and collaborating institutions, 
and acknowledge support from the 
DOE and NSF (USA);
CEA and CNRS/IN2P3 (France);
FASI, Rosatom and RFBR (Russia);
CNPq, FAPERJ, FAPESP and FUNDUNESP (Brazil);
DAE and DST (India);
Colciencias (Colombia);
CONACyT (Mexico);
KRF and KOSEF (Korea);
CONICET and UBACyT (Argentina);
FOM (The Netherlands);
STFC and the Royal Society (United Kingdom);
MSMT and GACR (Czech Republic);
CRC Program, CFI, NSERC and WestGrid Project (Canada);
BMBF and DFG (Germany);
SFI (Ireland);
The Swedish Research Council (Sweden);
CAS and CNSF (China);
and the
Alexander von Humboldt Foundation (Germany).

\clearpage

\section{Tables of Results}
Tables \ref{tab:fulldphi25} to \ref{tab:fullyb45} contain the normalized differential cross sections, $1/\sigma(\z) \times d\sigma(\z+\text{jet})/d X$, for the three angular variables \dphi, \adrap\ and \ayboost. 
For each bin we present the bin edges, bin average (defined in the main text), measured value, statistical uncertainty, uncorrelated systematic uncertainty, and the systematic uncertainties from the following sources, which are correlated across all bins and all distributions: 
\begin{enumerate}
\item $\Delta \phi, \Delta y, y_{\text{boost}}$ re-weighting; 
\item Jet y re-weighting; 
\item \Z\ p$_{\text{T}}$\ re-weighting; 
\item Jet p$_{\text{T}}$\ re-weighting; 
\item Data overlay on simulation; 
\item Muon 1/p$_{\text{T}}$\ resolution; 
\item Muon identification efficiency; 
\item Jet identification efficiency;
\item Jet energy resolution; 
\item Jet energy scale.
\end{enumerate}

\begin{table*}[h]
\caption{\label{tab:fulldphi25}The measured cross section in bins of \dphi\ for \Z+jet+$X$\ events with \ptz$>25$~GeV, normalized to the measured \z\ cross section.}
\begin{ruledtabular}
\tiny{\begin{tabular}{r@{.}l@{-}r@{.}lr@{.}lr@{.}lr@{.}lr@{.}lr@{.}lr@{.}lr@{.}lr@{.}lr@{.}lr@{.}lr@{.}lr@{.}lr@{.}lr@{.}lr@{.}lr@{.}lr@{.}lr@{.}lr@{.}lr@{.}lr@{.}lr@{.}lr@{.}lr@{.}lr@{.}lr@{.}lr@{.}lr@{.}lr@{.}lr@{.}lr@{.}lr@{.}lr@{.}lr@{.}lr@{.}lr@{.}lr@{.}lr@{.}lr@{.}lr@{.}lr@{.}lr@{.}lr@{.}lr@{.}l} 
\multicolumn{4}{c} \mbox{$\Delta\phi$~~~} & \multicolumn{2}{c} \mbox{$\langle \Delta\phi\rangle$~} &\multicolumn{2}{c} \mbox{ result } & \multicolumn{2}{c} \mbox{stat.~~~} & \multicolumn{2}{c} \mbox{uncorr.} & \multicolumn{4}{c} \mbox{source 1~} & \multicolumn{4}{c} \mbox{source 2~~} & \multicolumn{4}{c} \mbox{source 3~} & \multicolumn{4}{c} \mbox{source 4~} & \multicolumn{4}{c} \mbox{source 5~} & \multicolumn{4}{c} \mbox{source 6~} & \multicolumn{4}{c} \mbox{source 7~} & \multicolumn{4}{c} \mbox{source 8~} & \multicolumn{4}{c} \mbox{source 9~} & \multicolumn{4}{c} \mbox{source 10~~~} \\ 
\multicolumn{4}{c} \mbox{(rad)~} & \multicolumn{2}{c} \mbox{(rad)~~~} & \multicolumn{2}{c} \mbox{(1/rad)} & \multicolumn{2}{c} \mbox{unc. (\%)} & \multicolumn{2}{c} \mbox{unc. (\%)} &  \multicolumn{4}{c} \mbox{(\%)~~~~} & \multicolumn{4}{c} \mbox{(\%)~~~~~} & \multicolumn{4}{c} \mbox{(\%)~~~~} & \multicolumn{4}{c} \mbox{(\%)~~~~} & \multicolumn{4}{c} \mbox{(\%)~~~~} & \multicolumn{4}{c} \mbox{(\%)~~~~} & \multicolumn{4}{c} \mbox{(\%)~~~~} & \multicolumn{4}{c} \mbox{(\%)~~~~} & \multicolumn{4}{c} \mbox{(\%)~~~~} & \multicolumn{4}{c} \mbox{(\%)~~~~~~~~} \\ 
0&0 & 1&5 & ~~1&09 & 0&000282 & $\pm$12& & $ ~~\pm$4&0 &   16& &  -20& & 0&7 & -0&7 & -1&9 &   1&9 & -1&0 &   1&0 &  8&0 &  -8&0 & -3&2 &   1&3 & -3&8 &   5&3 &  2&4 &  -2&4 &-0&3 &  0&1 & -4&4 &   6&3\\ 
1&5 & 2&2 & ~~1&95 & 0&00422 & $\pm$6&9 & $ ~~\pm$0&8 &   3&2 &  -4&2 & 0&7 & -0&7 &-1&0 &  1&0 &-0&5 &  0&5 &  2&1 &  -2&1 & -3&5 &   4&2 & -4&4 &   3&4 &  1&3 &  -1&3 &-0&8 & -0&7 & -3&7 &   2&9\\ 
2&2 & 2&5 & ~~2&38 & 0&0193 & $\pm$5&5 & $ ~~\pm$0&9 &   1&4 &  -1&6 & 0&5 & -0&5 &-0&6 &  0&6 &-0&3 &  0&3 & 0&5 & -0&5 & -2&6 &   1&6 & -3&7 &   3&3 &0&1 & -0&1 &-0&3 & -0&2 & -2&8 &   2&4\\ 
2&5 & 2&7 & ~~2&61 & 0&0527 & $\pm$4&2 & $ ~~\pm$0&8 &   1&1 &  -1&4 & 0&5 & -0&5 &-0&3 &  0&3 &-0&2 &  0&2 & 0&6 & -0&6 & -1&7 &   1&9 & -2&4 &   2&3 & -1&2 &   1&2 &-0&1 & -0&3 & -2&9 &   2&8\\ 
2&7 & 2&9 & ~~2&81 & 0&113 & $\pm$2&8 & $ ~~\pm$0&6 &  0&8 &  -2&1 & 0&5 & -0&5 &-0&3 &  0&3 &-0&1 & 0&1 & 0&8 & -0&8 & -1&2 &   1&1 & -1&7 &   1&8 &-0&9 &  0&8 &0&1 &  0&2 & -1&7 &   2&1\\ 
2&9 & 3&2 & ~~3&04 & 0&332 & $\pm$1&7 & $ ~~\pm$0&4 & -0&3 &  0&3 & 0&2 & -0&2 &-0&5 &  0&5 &0&0 & -0&0 &-0&4 &  0&4 &  2&2 &  -1&8 & -1&0 &   1&2 & -1&1 &   1&1 & 0&2 & 0&0 & -1&3 &   1&3\\ 
\end{tabular}}
\end{ruledtabular}
\end{table*}

\begin{table*}[!t]
\caption{\label{tab:fulldrap25}The measured cross section in bins of \adrap\ for \Z+jet+$X$\ events with \ptz$>25$~GeV, normalized to the measured \z\ cross section.}
\begin{ruledtabular}
\tiny{\begin{tabular}{r@{.}l@{-}r@{.}lr@{.}lr@{.}lr@{.}lr@{.}lr@{.}lr@{.}lr@{.}lr@{.}lr@{.}lr@{.}lr@{.}lr@{.}lr@{.}lr@{.}lr@{.}lr@{.}lr@{.}lr@{.}lr@{.}lr@{.}lr@{.}lr@{.}lr@{.}lr@{.}lr@{.}lr@{.}lr@{.}lr@{.}lr@{.}lr@{.}lr@{.}lr@{.}lr@{.}lr@{.}lr@{.}lr@{.}lr@{.}lr@{.}lr@{.}lr@{.}lr@{.}lr@{.}lr@{.}lr@{.}l} 
\multicolumn{4}{c} \mbox{$|\Delta y|$~~~} & \multicolumn{2}{c} \mbox{$\langle |\Delta y|\rangle$} &\multicolumn{2}{c} \mbox{ result } & \multicolumn{2}{c} \mbox{stat.~~~.} & \multicolumn{2}{c} \mbox{uncorr.} & \multicolumn{4}{c} \mbox{source 1} & \multicolumn{4}{c} \mbox{source 2} & \multicolumn{4}{c} \mbox{source 3} & \multicolumn{4}{c} \mbox{source 4} & \multicolumn{4}{c} \mbox{source 5} & \multicolumn{4}{c} \mbox{source 6} & \multicolumn{4}{c} \mbox{source 7} & \multicolumn{4}{c} \mbox{source 8} & \multicolumn{4}{c} \mbox{source 9} & \multicolumn{4}{c} \mbox{source 10} \\ 
\multicolumn{4}{c} \mbox{~~ }  & \multicolumn{2}{c} \mbox{~~} & \multicolumn{2}{c} \mbox{~~} & \multicolumn{2}{c} \mbox{unc. (\%)} & \multicolumn{2}{c} \mbox{unc. (\%)} & \multicolumn{4}{c} \mbox{(\%)~~~~} & \multicolumn{4}{c} \mbox{(\%)~~~~} & \multicolumn{4}{c} \mbox{(\%)~~~~} & \multicolumn{4}{c} \mbox{(\%)~~~~} & \multicolumn{4}{c} \mbox{(\%)~~~~} & \multicolumn{4}{c} \mbox{(\%)~~~~} & \multicolumn{4}{c} \mbox{(\%)~~~~} & \multicolumn{4}{c} \mbox{(\%)~~~~} & \multicolumn{4}{c} \mbox{(\%)~~~~} & \multicolumn{4}{c} \mbox{(\%)~~~~} \\ 
0&00 & 0&40 & ~~~0&21 & 0&0791 & ~$\pm$2&6 & ~~$ \pm$0&6 &  0&5 & -0&5 & 0&3 & -0&3 &-0&4 &  0&4 &-0&1 & 0&1 &0&1 & -0&1 & 0&8 & -0&3 & -1&6 &   1&8 & -1&1 &   1&1 & 0&3 &  0&2 & -1&4 &   1&6\\ 
0&40 & 0&80 & ~~~0&61 & 0&0679 & ~$\pm$2&8 & ~~$ \pm$0&6 &  0&1 & -0&1 & 0&3 & -0&3 &-0&4 &  0&4 &-0&1 & 0&1 &-0&6 &  0&6 & 0&4 & -0&4 & -1&5 &   1&6 &-0&9 &  0&9 &-0&3 & -0&2 & -1&6 &   1&5\\ 
0&80 & 1&20 & ~~~1&02 & 0&0568 & ~$\pm$3&0 & ~~$ \pm$0&7 & -0&1 & 0&1 & 0&4 & -0&4 &-0&4 &  0&4 &-0&1 &  0&1 &-0&0 & 0&0 & 0&5 & -0&5 & -1&4 &   1&6 & -1&1 &   1&1 &0&1 & 0&1 & -1&5 &   1&6\\ 
1&20 & 1&55 & ~~~1&37 & 0&0452 & ~$\pm$3&6 & ~~$ \pm$0&9 & -0&2 &  0&2 & 0&2 & -0&2 &-0&4 &  0&4 &-0&1 &  0&1 & 0&1 & -0&1 & 0&3 & -0&8 & -1&6 &   1&4 &-0&3 &  0&3 &-0&1 & -0&6 & -1&8 &   1&6\\ 
1&55 & 2&05 & ~~~1&78 & 0&0274 & ~$\pm$3&8 & ~~$ \pm$0&9 & -0&4 &  0&4 &-0&2 &  0&2 &-0&4 &  0&4 &-0&1 &  0&1 &  1&1 &  -1&1 &  1&1 & -0&1 & -1&3 &   1&8 & -1&2 &   1&2 & 0&4 &  0&4 & -1&8 &   2&5\\ 
2&05 & 4&50 & ~~~2&89 & 0&00480 &~$\pm$4&0 & ~~$ \pm$1&1 & -0&6 &  0&6 & -1&1 &   1&1 &-0&5 &  0&5 &-0&1 & 0&1 &  1&3 &  -1&3 &-0&1 & -0&8 & -2&0 &   1&5 &-0&5 &  0&5 &-0&0 & -0&1 & -3&0 &   2&6\\ 
\end{tabular}}
\end{ruledtabular}
\end{table*}

\begin{table*}[!htb]
\caption{\label{tab:fullyb25}The measured cross section in bins of \ayb\ ($|y_b|$) for \Z+jet+$X$\ events with \ptz$>25$~GeV, normalized to the measured \z\ cross section.}
\begin{ruledtabular}
\tiny{\begin{tabular}{r@{.}l@{-}r@{.}lr@{.}lr@{.}lr@{.}lr@{.}lr@{.}lr@{.}lr@{.}lr@{.}lr@{.}lr@{.}lr@{.}lr@{.}lr@{.}lr@{.}lr@{.}lr@{.}lr@{.}lr@{.}lr@{.}lr@{.}lr@{.}lr@{.}lr@{.}lr@{.}lr@{.}lr@{.}lr@{.}lr@{.}lr@{.}lr@{.}lr@{.}lr@{.}lr@{.}lr@{.}lr@{.}lr@{.}lr@{.}lr@{.}lr@{.}lr@{.}lr@{.}lr@{.}lr@{.}lr@{.}l} 
\multicolumn{4}{c} \mbox{$|y_b|$~~~} & \multicolumn{2}{c} \mbox{$\langle |y_b|\rangle$} &\multicolumn{2}{c} \mbox{ result } & \multicolumn{2}{c} \mbox{stat.~~~} & \multicolumn{2}{c} \mbox{uncorr.} & \multicolumn{4}{c} \mbox{source 1} & \multicolumn{4}{c} \mbox{source 2} & \multicolumn{4}{c} \mbox{source 3} & \multicolumn{4}{c} \mbox{source 4} & \multicolumn{4}{c} \mbox{source 5} & \multicolumn{4}{c} \mbox{source 6} & \multicolumn{4}{c} \mbox{source 7} & \multicolumn{4}{c} \mbox{source 8} & \multicolumn{4}{c} \mbox{source 9} & \multicolumn{4}{c} \mbox{source 10} \\ 
\multicolumn{4}{c} \mbox{ }  & \multicolumn{2}{c} \mbox{} & \multicolumn{2}{c} \mbox{} & \multicolumn{2}{c} \mbox{unc. (\%)} & \multicolumn{2}{c} \mbox{unc. (\%)} & \multicolumn{4}{c} \mbox{(\%)~~~~} & \multicolumn{4}{c} \mbox{(\%)~~~~} & \multicolumn{4}{c} \mbox{(\%)~~~~} & \multicolumn{4}{c} \mbox{(\%)~~~~} & \multicolumn{4}{c} \mbox{(\%)~~~~} & \multicolumn{4}{c} \mbox{(\%)~~~~} & \multicolumn{4}{c} \mbox{(\%)~~~~} & \multicolumn{4}{c} \mbox{(\%)~~~~} & \multicolumn{4}{c} \mbox{(\%)~~~~} & \multicolumn{4}{c} \mbox{(\%)~~~~} \\ 
0&00 & 0&20 & ~~~0&11 & ~0&124 & ~$\pm$2&9 & ~$ \pm$0&6 &   1&0 &  -1&0 & 0&4 & -0&4 &-0&4 &  0&4 &-0&2 &  0&2 &-0&5 &  0&5 & 0&2 & -0&9 & -1&8 &   1&6 & -1&0 &   1&0 &-0&1 & -0&2 & -1&3 &   1&4\\ 
0&20 & 0&40 & ~~~0&31 & ~0&115 & ~$\pm$3&0 & ~$ \pm$0&7 &  0&8 & -0&8 & 0&5 & -0&5 &-0&4 &  0&4 &-0&1 &  0&1 &-0&3 &  0&3 &  1&0 & 0&1 & -1&3 &   1&5 & -1&0 &   1&0 & 0&1 &  0&1 & -1&2 &   1&5\\ 
0&40 & 0&60 & ~~~0&51 & ~0&105 & ~$\pm$3&2 & ~$ \pm$0&7 &  0&8 & -0&8 & 0&4 & -0&4 &-0&4 &  0&4 &-0&1 & 0&1 & 0&3 & -0&3 & 0&4 & -0&8 & -1&6 &   1&7 &-0&7 &  0&7 &0&1 & -0&3 & -1&2 &   1&4\\ 
0&60 & 0&80 & ~~~0&70 & ~0&0895 & ~$\pm$3&4 & ~$ \pm$0&8 &   1&1 &  -1&1 & 0&5 & -0&5 &-0&4 &  0&4 &-0&1 &  0&1 & 0&6 & -0&6 & 0&5 & -0&1 & -1&3 &   1&7 &-0&9 &  0&9 & 0&2 &  0&4 & -2&1 &   2&0\\ 
0&80 & 1&00 & ~~~0&91 & ~0&0701 & ~$\pm$3&9 & ~$ \pm$0&9 &  0&8 & -0&8 & 0&3 & -0&3 &-0&4 &  0&4 &-0&1 & 0&1 & 0&5 & -0&5 & 0&9 & -0&5 & -1&4 &   2&1 & -1&1 &   1&1 & 0&2 & -0&1 & -1&8 &   2&6\\ 
1&00 & 1&25 & ~~~1&13 & ~0&0442 & ~$\pm$4&4 & ~$ \pm$1&0 &  0&9 & -0&9 & 0&2 & -0&2 &-0&4 &  0&4 &-0&1 & 0&1 & 0&4 & -0&4 & 0&7 & -0&4 & -1&6 &   1&4 & -1&1 &   1&1 &-0&1 &  0&2 & -2&4 &   2&1\\ 
1&25 & 2&25 & ~~~1&62 & ~0&0115 & ~$\pm$4&2 & ~$ \pm$1&1 &  0&3 & -0&3 &-0&7 &  0&7 &-0&5 &  0&5 &-0&0 & 0&0 & 0&6 & -0&6 & 0&2 & -0&2 & -1&8 &   1&5 &-0&5 &  0&5 &0&1 & -0&3 & -3&1 &   2&2\\ 
\end{tabular}}
\end{ruledtabular}
\end{table*}

\begin{table*}[!htb]
\caption{\label{tab:fulldphi45}The measured cross section in bins of \dphi\ for \Z+jet+$X$\ events with \ptz$>45$~GeV, normalized to the measured \z\ cross section.}
\begin{ruledtabular}
\tiny{\begin{tabular}{r@{.}l@{-}r@{.}lr@{.}lr@{.}lr@{.}lr@{.}lr@{.}lr@{.}lr@{.}lr@{.}lr@{.}lr@{.}lr@{.}lr@{.}lr@{.}lr@{.}lr@{.}lr@{.}lr@{.}lr@{.}lr@{.}lr@{.}lr@{.}lr@{.}lr@{.}lr@{.}lr@{.}lr@{.}lr@{.}lr@{.}lr@{.}lr@{.}lr@{.}lr@{.}lr@{.}lr@{.}lr@{.}lr@{.}lr@{.}lr@{.}lr@{.}lr@{.}lr@{.}lr@{.}lr@{.}lr@{.}l} 
\multicolumn{4}{c} \mbox{$\Delta\phi$~~~} & \multicolumn{2}{c} \mbox{$\langle \Delta\phi\rangle$~} &\multicolumn{2}{c} \mbox{ result } & \multicolumn{2}{c} \mbox{stat.~~~} & \multicolumn{2}{c} \mbox{uncorr.} & \multicolumn{4}{c} \mbox{source 1~} & \multicolumn{4}{c} \mbox{source 2~~} & \multicolumn{4}{c} \mbox{source 3~} & \multicolumn{4}{c} \mbox{source 4~} & \multicolumn{4}{c} \mbox{source 5~} & \multicolumn{4}{c} \mbox{source 6~} & \multicolumn{4}{c} \mbox{source 7~} & \multicolumn{4}{c} \mbox{source 8~} & \multicolumn{4}{c} \mbox{source 9~} & \multicolumn{4}{c} \mbox{source 10~~~} \\ 
\multicolumn{4}{c} \mbox{(rad)~} & \multicolumn{2}{c} \mbox{(rad)~~~} & \multicolumn{2}{c} \mbox{(1/rad)} & \multicolumn{2}{c} \mbox{unc. (\%)} & \multicolumn{2}{c} \mbox{unc. (\%)} &  \multicolumn{4}{c} \mbox{(\%)~~~~} & \multicolumn{4}{c} \mbox{(\%)~~~~~} & \multicolumn{4}{c} \mbox{(\%)~~~~} & \multicolumn{4}{c} \mbox{(\%)~~~~} & \multicolumn{4}{c} \mbox{(\%)~~~~} & \multicolumn{4}{c} \mbox{(\%)~~~~} & \multicolumn{4}{c} \mbox{(\%)~~~~} & \multicolumn{4}{c} \mbox{(\%)~~~~} & \multicolumn{4}{c} \mbox{(\%)~~~~} & \multicolumn{4}{c} \mbox{(\%)~~~~~~~~} \\ 
0&0 & 1&5 & ~~1&09 & 0&0000488 & $\pm$26& & $ \pm$12& &   25& &  -30& & 0&8 & -0&8 &  2&5 &  -2&5 & -2&5 &   2&5 &  8&1 &  -8&1 & -6&0 &   12& & -6&8 &   5&9 &  7&2 &  -7&2 & -4&4 &  -2&1 & -2&8 &  -1&5\\ 
1&5 & 2&2 & ~~1&95 & 0&000761 & $\pm$16& & $ \pm$1&1 &   4&7 &  -6&1 & 0&7 & -0&7 &  1&5 &  -1&5 & -1&5 &   1&5 &  1&1 &  -1&1 & -2&7 &   3&5 & -5&9 &   5&6 &  1&9 &  -1&9 &-0&8 &  -1&1 & -1&0 & 0&1\\ 
2&2 & 2&5 & ~~2&38 & 0&00657 & $\pm$9&1 & $ \pm$1&2 &   1&9 &  -2&1 & 0&8 & -0&8 &  1&2 &  -1&2 &-0&1 &  0&1 &  1&4 &  -1&4 & -3&0 &   2&3 & -4&0 &   3&9 &-0&2 &  0&2 & 0&1 & 0&0 &-0&1 &  0&7\\ 
2&5 & 2&7 & ~~2&61 & 0&0191 & $\pm$6&9 & $ \pm$1&1 &   1&2 &  -1&5 & 0&3 & -0&3 &  1&1 &  -1&1 &-0&5 &  0&5 & -1&7 &   1&7 & -2&6 &  0&9 & -3&3 &   3&3 &-0&7 &  0&7 &-0&3 & -0&2 & -1&2 &  0&8\\ 
2&7 & 2&9 & ~~2&81 & 0&0383 & $\pm$4&8 & $ \pm$0&8 &  0&9 &  -2&3 & 0&5 & -0&5 &  1&0 &  -1&0 &-0&5 &  0&5 & 0&5 & -0&5 & -2&1 &   2&1 & -2&6 &   2&4 &-0&5 &  0&5 &-0&2 & 0&0 &-0&8 &  0&5\\ 
2&9 & 3&2 & ~~3&04 & 0&135 & $\pm$2&5 & $ \pm$0&5 & -0&4 &  0&4 & 0&3 & -0&3 & 0&7 & -0&7 &-0&7 &  0&7 &-0&0 & 0&0 &  1&2 &  -1&3 & -1&4 &   1&5 &-0&9 &  0&9 &0&1 & 0&5 &-0&3 &  0&3\\ 
\end{tabular}}
\end{ruledtabular}
\end{table*}

\begin{table*}[!htb]
\caption{\label{tab:fulldrap45}The measured cross section in bins of \adrap\ for \Z+jet+$X$\ events with \ptz$>45$~GeV, normalized to the measured \z\ cross section.}
\begin{ruledtabular}
\tiny{\begin{tabular}{r@{.}l@{-}r@{.}lr@{.}lr@{.}lr@{.}lr@{.}lr@{.}lr@{.}lr@{.}lr@{.}lr@{.}lr@{.}lr@{.}lr@{.}lr@{.}lr@{.}lr@{.}lr@{.}lr@{.}lr@{.}lr@{.}lr@{.}lr@{.}lr@{.}lr@{.}lr@{.}lr@{.}lr@{.}lr@{.}lr@{.}lr@{.}lr@{.}lr@{.}lr@{.}lr@{.}lr@{.}lr@{.}lr@{.}lr@{.}lr@{.}lr@{.}lr@{.}lr@{.}lr@{.}lr@{.}lr@{.}l} 
\multicolumn{4}{c} \mbox{$|\Delta y|$~~~} & \multicolumn{2}{c} \mbox{$\langle |\Delta y|\rangle$} &\multicolumn{2}{c} \mbox{ result } & \multicolumn{2}{c} \mbox{stat.~~~.} & \multicolumn{2}{c} \mbox{uncorr.} & \multicolumn{4}{c} \mbox{source 1} & \multicolumn{4}{c} \mbox{source 2} & \multicolumn{4}{c} \mbox{source 3} & \multicolumn{4}{c} \mbox{source 4} & \multicolumn{4}{c} \mbox{source 5} & \multicolumn{4}{c} \mbox{source 6} & \multicolumn{4}{c} \mbox{source 7} & \multicolumn{4}{c} \mbox{source 8} & \multicolumn{4}{c} \mbox{source 9} & \multicolumn{4}{c} \mbox{source 10} \\ 
\multicolumn{4}{c} \mbox{~~ }  & \multicolumn{2}{c} \mbox{~~} & \multicolumn{2}{c} \mbox{~~} & \multicolumn{2}{c} \mbox{unc. (\%)} & \multicolumn{2}{c} \mbox{unc. (\%)} & \multicolumn{4}{c} \mbox{(\%)~~~~} & \multicolumn{4}{c} \mbox{(\%)~~~~} & \multicolumn{4}{c} \mbox{(\%)~~~~} & \multicolumn{4}{c} \mbox{(\%)~~~~} & \multicolumn{4}{c} \mbox{(\%)~~~~} & \multicolumn{4}{c} \mbox{(\%)~~~~} & \multicolumn{4}{c} \mbox{(\%)~~~~} & \multicolumn{4}{c} \mbox{(\%)~~~~} & \multicolumn{4}{c} \mbox{(\%)~~~~} & \multicolumn{4}{c} \mbox{(\%)~~~~~} \\ 
0&00 & 0&40 & ~~~0&21 & 0&0331 & ~$\pm$3&9 & ~$ \pm$0&8 &   0&0 &   0&0 & 0&3 & -0&3 & 0&7 & -0&7 &-0&7 &  0&7 & 0&3 & -0&3 & 0&4 & -0&2 & -1&9 &   1&8 &-0&7 &  0&7 &-0&0 & -0&1 &-0&5 &  0&3\\ 
0&40 & 0&80 & ~~~0&61 & 0&0291 & ~$\pm$4&1 & ~$ \pm$0&8 &   0&0 &   0&0 & 0&3 & -0&3 & 0&7 & -0&7 &-0&7 &  0&7 &-0&7 &  0&7 &-0&1 & -0&1 & -1&9 &   2&0 &-0&6 &  0&6 &-0&1 & -0&0 &-0&4 &  0&4\\ 
0&80 & 1&20 & ~~~1&02 & 0&0214 & ~$\pm$4&7 & ~$ \pm$0&9 &   0&0 &   0&0 & 0&4 & -0&4 & 0&8 & -0&8 &-0&8 &  0&8 &-0&0 & 0&0 &-0&2 & -0&2 & -1&9 &   1&7 &-0&9 &  0&9 &-0&1 & -0&2 &-0&9 &  0&3\\ 
1&20 & 1&55 & ~~~1&37 & 0&0156 & ~$\pm$5&9 & ~$ \pm$1&3 &   0&0 &   0&0 & 0&4 & -0&4 & 0&8 & -0&8 &-0&8 &  0&8 & 0&3 & -0&3 &0&1 & -0&3 & -1&8 &   1&8 &-0&4 &  0&4 &0&1 & 0&0 &-0&2 &  0&5\\ 
1&55 & 2&05 & ~~~1&78 & 0&00960 & ~$\pm$6&2 & ~$ \pm$1&3 &   0&0 &   0&0 &-0&0 & 0&0 & 0&9 & -0&9 &-0&8 &  0&8 & 0&1 & -0&1 & 0&4 & -0&7 & -1&9 &   2&5 &-0&9 &  0&9 & 0&2 &  0&5 &-0&3 &   1&2\\ 
2&05 & 4&50 & ~~~2&89 & 0&00127 & ~$\pm$7&6 & ~$ \pm$1&6 &   0&0 &   0&0 & -1&1 &   1&1 &  1&0 &  -1&0 &-0&6 &  0&6 &  1&2 &  -1&2 & -1&2 & -0&0 & -2&4 &   2&1 &-0&2 &  0&2 &-0&2 & 0&1 &-0&7 &  0&2\\ 
\end{tabular}}
\end{ruledtabular}
\end{table*}

\begin{table*}[!htb]
\caption{\label{tab:fullyb45}The measured cross section in bins of \ayb\ ($|y_b|$) for \Z+jet+$X$\ events with \ptz$>45$~GeV, normalized to the measured \z\ cross section.}
\begin{ruledtabular}
\tiny{\begin{tabular}{r@{.}l@{-}r@{.}lr@{.}lr@{.}lr@{.}lr@{.}lr@{.}lr@{.}lr@{.}lr@{.}lr@{.}lr@{.}lr@{.}lr@{.}lr@{.}lr@{.}lr@{.}lr@{.}lr@{.}lr@{.}lr@{.}lr@{.}lr@{.}lr@{.}lr@{.}lr@{.}lr@{.}lr@{.}lr@{.}lr@{.}lr@{.}lr@{.}lr@{.}lr@{.}lr@{.}lr@{.}lr@{.}lr@{.}lr@{.}lr@{.}lr@{.}lr@{.}lr@{.}lr@{.}lr@{.}lr@{.}l} 
\multicolumn{4}{c} \mbox{$|y_b|$~~~} & \multicolumn{2}{c} \mbox{$\langle |y_b|\rangle$} &\multicolumn{2}{c} \mbox{ result } & \multicolumn{2}{c} \mbox{stat.~~~} & \multicolumn{2}{c} \mbox{uncorr.} & \multicolumn{4}{c} \mbox{source 1} & \multicolumn{4}{c} \mbox{source 2} & \multicolumn{4}{c} \mbox{source 3} & \multicolumn{4}{c} \mbox{source 4} & \multicolumn{4}{c} \mbox{source 5} & \multicolumn{4}{c} \mbox{source 6} & \multicolumn{4}{c} \mbox{source 7} & \multicolumn{4}{c} \mbox{source 8} & \multicolumn{4}{c} \mbox{source 9} & \multicolumn{4}{c} \mbox{source 10} \\ 
\multicolumn{4}{c} \mbox{ }  & \multicolumn{2}{c} \mbox{} & \multicolumn{2}{c} \mbox{} & \multicolumn{2}{c} \mbox{unc. (\%)} & \multicolumn{2}{c} \mbox{unc. (\%)} & \multicolumn{4}{c} \mbox{(\%)~~~~} & \multicolumn{4}{c} \mbox{(\%)~~~~} & \multicolumn{4}{c} \mbox{(\%)~~~~} & \multicolumn{4}{c} \mbox{(\%)~~~~} & \multicolumn{4}{c} \mbox{(\%)~~~~} & \multicolumn{4}{c} \mbox{(\%)~~~~} & \multicolumn{4}{c} \mbox{(\%)~~~~} & \multicolumn{4}{c} \mbox{(\%)~~~~} & \multicolumn{4}{c} \mbox{(\%)~~~~} & \multicolumn{4}{c} \mbox{(\%)~~~~} \\ 
0&00 & 0&20 & ~~~0&11 & 0&0523 & ~~$\pm$4&3 & ~~~$\pm$0&9 &  0&8 & -0&8 & 0&3 & -0&3 & 0&7 & -0&7 &-0&9 &  0&9 &-0&8 &  0&8 &-0&2 & -0&7 & -2&1 &   2&0 &-0&7 &  0&7 &-0&0 & -0&1 &-0&2 &  0&1\\ 
0&20 & 0&40 & ~~~0&31 & 0&0450 & ~~$\pm$4&6 & ~~~$\pm$0&9 &  0&7 & -0&7 & 0&4 & -0&4 & 0&7 & -0&7 &-0&8 &  0&8 & 0&6 & -0&6 & 1&0 & -0&3 & -1&8 &   1&8 &-0&8 &  0&8 &0&1 &  0&2 &-0&3 &  0&4\\ 
0&40 & 0&60 & ~~~0&51 & 0&0436 & ~~$\pm$4&7 & ~~~$\pm$1&0 &  0&5 & -0&5 & 0&3 & -0&3 & 0&7 & -0&7 &-0&7 &  0&7 &-0&6 &  0&6 & 0&3 & -0&0 & -2&1 &   2&0 &-0&4 &  0&4 &-0&1 & -0&2 &-0&4 &  0&3\\ 
0&60 & 0&80 & ~~~0&70 & 0&0350 & ~~$\pm$5&3 & ~~~$\pm$1&1 &  0&6 & -0&6 & 0&2 & -0&2 & 0&7 & -0&7 &-0&7 &  0&7 &-0&5 &  0&5 &-0&1 & -0&2 & -1&8 &   2&0 & -1&0 &   1&0 & 0&1 &  0&4 &-0&3 &  0&8\\ 
0&80 & 1&00 & ~~~0&91 & 0&0257 & ~~$\pm$6&1 & ~~~$\pm$1&2 &  0&2 & -0&2 & 0&2 & -0&2 & 0&9 & -0&9 &-0&6 &  0&6 & 0&4 & -0&4 &-0&1 & -0&5 & -1&9 &   2&3 &-0&6 &  0&6 &0&1 & -0&2 &-0&6 &  0&6\\ 
1&00 & 1&25 & ~~~1&13 & 0&0155 & ~~$\pm$7&0 & ~~~$\pm$1&4 & 0&1 & -0&1 &-0&0 & 0&0 & 0&9 & -0&9 &-0&5 &  0&5 &  1&0 &  -1&0 &-0&2 & -0&5 & -1&8 &   1&8 & -1&0 &   1&0 &-0&4 & -0&0 &-0&8 &  0&6\\ 
1&25 & 2&25 & ~~~1&62 & 0&00272 & ~~$\pm$8&1 & ~~~$\pm$1&5 & -0&7 &  0&7 &-0&2 &  0&2 &  1&5 &  -1&5 &-0&5 &  0&5 &  1&5 &  -1&5 & -1&9 &   1&4 & -1&9 &   1&6 &-0&3 &  0&3 &-0&1 & -0&4 & -1&8 &  0&2\\ 
\end{tabular}}
\end{ruledtabular}
\end{table*}


\begin{thebibliography}{99}
\bibitem{rapidity}Rapidity is defined as $y = 0.5\ln[(E+p_z)/(E-p_z)]$, where $E$\ is the energy, and $p_z$\ is the component of momentum parallel to the proton beam direction.
\bibitem{eta} $\eta = -\ln [\tan(\theta/2)]$, where $\theta$\ is the polar angle, defined with respect to the proton beam direction.
\end{thebibliography}
\end{document}